\documentstyle[e-e-ijmpa,twoside, epsf]{article}
\input psfig

\renewcommand{\thefootnote}{\fnsymbol{footnote}}      

\begin{document}

\begin{flushright}
SLAC--PUB--8349\\
January 2000
\end{flushright}
\bigskip\bigskip

\normalsize\textlineskip
\pagestyle{empty}

\def\grtsim{\,\,\rlap{\raise 3pt\hbox{$>$}}{\lower 3pt\hbox{$\sim$}}\,\,}
\def\lsim{\,\,\rlap{\raise 3pt\hbox{$<$}}{\lower 3pt\hbox{$\sim$}}\,\,}

\newcommand{\gege}{\gamma \, e \to \gamma \, e}
\newcommand{\gggg}{\gamma \gamma \to \gamma \gamma}

\title{SIGNATURES OF EXTRA DIMENSIONS AT  \boldmath$e \gamma$ AND \boldmath$\gamma \gamma$ COLLIDERS}

\author{HOOMAN DAVOUDIASL%
\footnote{Work supported by the Department of Energy, 
Contract DE-AC03 76SF00515.}
\footnote{Invited talk at e-e- 99, 3rd International Workshop on Electron-Electron 
Interactions at TeV Energies, December 10-12, 1999 University of
California, Santa Cruz.} 
}

\address{Stanford Linear Accelerator Center \\
Stanford, CA 94309, USA}

\maketitle\abstracts{We study the processes $\gege$ and $\gggg$, in the context of the proposal for Weak Scale Quantum Gravity
(WSQG) with large extra dimensions.  With an ultraviolet cutoff $M_S \sim 1$ TeV for the effective gravity theory, the cross
sections obtained for these processes at the Next Linear Collider (NLC), with the $e \gamma$ an $\gamma \gamma$ options,
deviate from the predictions of the Standard Model significantly.  Our results suggest that, for typical proposed NLC energies and
luminosities, the predictions of WSQG can be tested in the range 1 TeV$\lsim M_S \lsim$ 10 TeV, making $e \gamma$ an $\gamma \gamma$ colliders
important tools for probing WSQG.}

\setcounter{footnote}{0}
\renewcommand{\thefootnote}{\alph{footnote}}

\vspace*{1pt}\textlineskip        
\section{Introduction}

The following is based on the talk with the same title delivered at \boldmath$e^- e^- 99$ by the author at the University of California, Santa Cruz.  Most of the results and the discussion presented here are taken from Refs.~[\citenum{HD1}] and [\citenum{HD2}].

The idea of using extra dimensions in describing physical phenomena is a fairly mature one and dates back to the early decades of the
twentieth century.  During that time, attempts at unifying the theories of electromagnetism and gravitation were made by assuming the
existence of an extra spatial dimension \cite{NKK}.  More recently, extra dimensions have been considered in the context of super
string theories.  A new application of extra dimensional theories has been proposed in 
Refs.~[\citenum{ADD1}] and [\citenum{ADD2}], where it
was suggested that the fundamental scale of gravity $M_F$ could be as low as the weak scale $\Lambda_w \sim 1$ TeV, assuming
that there were $n$ large compactified extra dimensions of size $R$.  Guass' law in $4 + n$ dimensions then yields
\begin{equation}
M_P^2 \sim M_F^{n + 2} R^n, 
\label{MP} 
\end{equation} 
where $M_P \sim 10^{19}$ GeV is the Planck mass.  The above relation (\ref{MP}) can be viewed as a reformulation of the 
hierarchy problem, in the sense that now one has the task of explaining the size of the extra dimensions.

It was shown in Refs.~[\citenum{ADD1}] and [\citenum{ADD2}] that gravitational data allow $n \geq 2$, and for $2 \leq n \leq 6$
relation (\ref{MP}) gives 1 fm $\lsim R \lsim$ 1 mm.  This proposal has significant phenomenological implications for collider
experiments at the scale $\Lambda_w$, where Weak Scale Quantum Gravity (WSQG) effects are assumed to become strong.  Lately,
a great deal of effort has been made to constrain the proposal for WSQG \cite{Recent,SNGAM}.  In the case of $n = 2$, the most
stringent constraints come from astrophysical and cosmological observations\cite{ADD2}, and it is argued that $M_F \grtsim 100$
TeV \cite{SNGAM}.  However, terrestrial experimental data have constrained WSQG to have $M_F \grtsim 1$ TeV, and in the case of
$n \geq 3$, there is no evidence for a more severe constraint.  Assuming that quantum gravity effects are important at a scale $M_S
\sim 1$ TeV implies that future colliders with center of mass energy $\sqrt{s} \sim M_S$ will be able to probe WSQG.  We will
assume that $M_S = M_F$ in the rest of our discussion, for simplicity.

One possible future collider is the Next Linear Collider (NLC) with $\sqrt{s} \sim 1$ TeV.  It has been shown\cite{Ginzburg} that it is
possible to obtain $\gamma$-beams with energy and luminosity comparable to those of the $e$-beams at such a facility, using
Compton back scattered laser beams.  Assuming the availability of such high energy $\gamma$-beams, we compute the cross
sections for the processes $\gege$ and $\gggg$\footnote{We note that the leading order contributions of WSQG to these processes,
presented here, have the same form as those obtained from a leading order string theoretic calculation, although the string theoretic
results have a different origin.\cite{CPP}} \, and will show that these processes can be used to probe WSQG over the
phenomenologically interesting range 1 TeV$\lsim M_S \lsim$ 10 TeV, in which the scale of physics related to the question of 
hierarchy is expected to lie.

\section{\boldmath$\gege$ at an \boldmath$e \gamma$ collider}

The Feynman diagrams that contribute to the process $\gege$ in the Standard Model (SM) at the leading order are the tree level
$s$- and $u$-channel diagrams.  The leading WSQG contribution results from a sum over a tower of Kaluza-Klein (KK) gravitons
exchanged in the $t$-channel.  This sum is divergent and is regulated here by using $M_S$ as an ultraviolet cutoff.  Since we 
do not know the fundamental theory of gravity, the WSQG contribution that is obtained in this way can in principle have 
an unknown coefficient $w$.\cite{JH}  Then, the total amplitude including the contributions of SM and WSQG is given by 
${\cal M}^{(TOT)} = {\cal M}_{_{SM}}^{(s)} + {\cal M}_{_{SM}}^{(u)} + w {\cal M}_{_{WSQG}}^{(t)}$.  
However, for an order of magnitude estimate of the size of the WSQG contribution, using ${\cal
M}^{(TOT)}$ with $w = \pm 1$, as we do later, is reasonable.  

Let ${\cal M}_{i j k l}$, $i, j, k, l = \pm$, denote the helicity amplitudes for $\gege$, where $(i, j)$ are the helicities of the initial state
$(\gamma, e)$, and $(k, l)$ are the helicities of the final state $(\gamma, e)$, respectively.  We define $|{\cal M}_{i j}|^2$ by 
\begin{equation}
|{\cal M}_{i j}|^2 \equiv \sum_{k, l} |{\cal M}_{i j k l}|^2, 
\label{Mij2} 
\end{equation} 
where the summation is performed over the final state helicities.  We find,
\begin{equation}
|{\cal M}^{(TOT)}_{+ j}|^2 = \frac{- 32 \pi^2}{s \, u} \left[\alpha + w \left(\frac{s \, u \, D_n}{2 \, M_S^4}\right)\right]^2 
\left[s^2 (1 + j) + u^2 (1 - j)\right],
\label{M+j2}
\end{equation}
where $D_n$ is given by \cite{Han}
\begin{equation}
D_n (x) \approx \ln \left(\frac{M_S^2}{|x|}\right) \, \, \, \,  {\rm for} \, \, 
\, \, n = 2 \, \, ; \, \, D_n (x) \approx \left(\frac{2}{n - 2}\right) \, \, \, \, {\rm for} \, \, \, \, n > 2.
\label{Dnx}
\end{equation}

Let $E_e$ be the electron beam energy, and $E_\gamma$ be the scattered $\gamma$ 
energy in the laboratory frame.  The fraction of
the beam energy taken away by the photon is then 
\begin{equation}
x = \frac{E_\gamma}{E_e}.
\label{x}
\end{equation}

We take the laser photons to have energy $E_l$.  Then, the maximum value of $x$ is given by $x_{max} = (z)/(1 + z)$, where $z = 4
E_e E_l/m_e^2$, and $m_e$ is the electron mass.  One cannot increase $x_{max}$ simply by increasing $E_l$, since this makes the
process less efficient because of $e^+ e^-$ pair production through the interactions of the laser photons and the backward
scattered $\gamma$-beam.  The optimal value for $z$ is given by $z_{_{OPT}} = 2 \left(1 + {\sqrt 2}\right)$.  The photon number
density $f(x, P_e, P_l)$ and average helicity $\xi_2 (x, P_e, P_l)$ are functions of $x$, $P_e$, $P_l$, and $z$, however, we always set
$z = z_{_{OPT}}$ in our calculations.  The expressions for these two functions can be found in Ref.~[\citenum{Ginzburg}].

For various choices of $(P_{e_1}, P_{l_1})$ of the $\gamma$-beam and
$P_{e_2}$ of the electron beam, the differential cross section $d \sigma/d \Omega$ is given by
\begin{equation}
\frac{d \sigma}{d \Omega} = \frac{1}{(8 \, \pi)^2} \int\frac{d x f(x)}{x \, s_{e
e}}
\left[\left(\frac{1 + P_{e_2}\, \xi_2(x)}{2}\right)|{\cal M}_{++}|^2 + \left(\frac{1 -   P_{e_2} \, 
\xi_2(x)}{2}\right)|{\cal M}_{+-}|^2\right],
\label{diffcs}
\end{equation}
where $s_{ee} = 4 E_e^2$.  Different choices of $(P_{e_1}, P_{l_1})$, in $(f(x),
 \xi_2(x))$, and $P_{e_2}$  yield different polarization cross
sections.  We take $|P_l| = 1$ and $|P_e| 
= 0.9$ for our calculations.  Note that the expressions for $|{\cal M}_{++}|^2$ and $|{\cal M}_{
+-}|^2$ are actually functions of the $\gamma \, e$ center of
mass energy squared ${\hat s} = x \, s_{ee}$, and the center of mass scattering 
angle $\theta_{cm}$.  We also have $t \to {\hat t}$ and $u \to
{\hat u}$, where ${\hat t} = - ({\hat s}/2 ) (1 - \cos \theta_{cm})$ and ${\hat 
u} = - ({\hat s}/2 ) (1 + \cos \theta_{cm})$.  We use Eq. (\ref{diffcs}) and the cuts 
$\theta_{cm} \in [\pi/6, 5 \pi/6] \, \, ; \, \, x \in [0.1, x_{max}]$ to compute the $\gege$ cross sections.  
To obtain the $M_S$ reach, we have used the $\chi^2 (M_S)$ variable given by 
\begin{equation} \chi^2 (M_S) =
\left(\frac{L}{\sigma_{_{SM}}}\right)\left[\sigma_{_{SM}} - \sigma (M_S)\right]^2, 
\label{chi2} 
\end{equation}
where $L$ is the luminosity, $\sigma_{_{SM}}$ is the SM cross section, and $\sigma (M_S)$ is the SM $\pm$ WSQG cross section
as a function of $M_S$.  We have taken $L = 100$ fb$^{-1}$ per year for our calculations.  We demand $\chi^2 (M_S) \geq 2.706$,
corresponding to a one-sided $95\%$ confidence level.

\begin{figure}[htbp] 
\centerline{\epsfxsize=10truecm \epsfbox{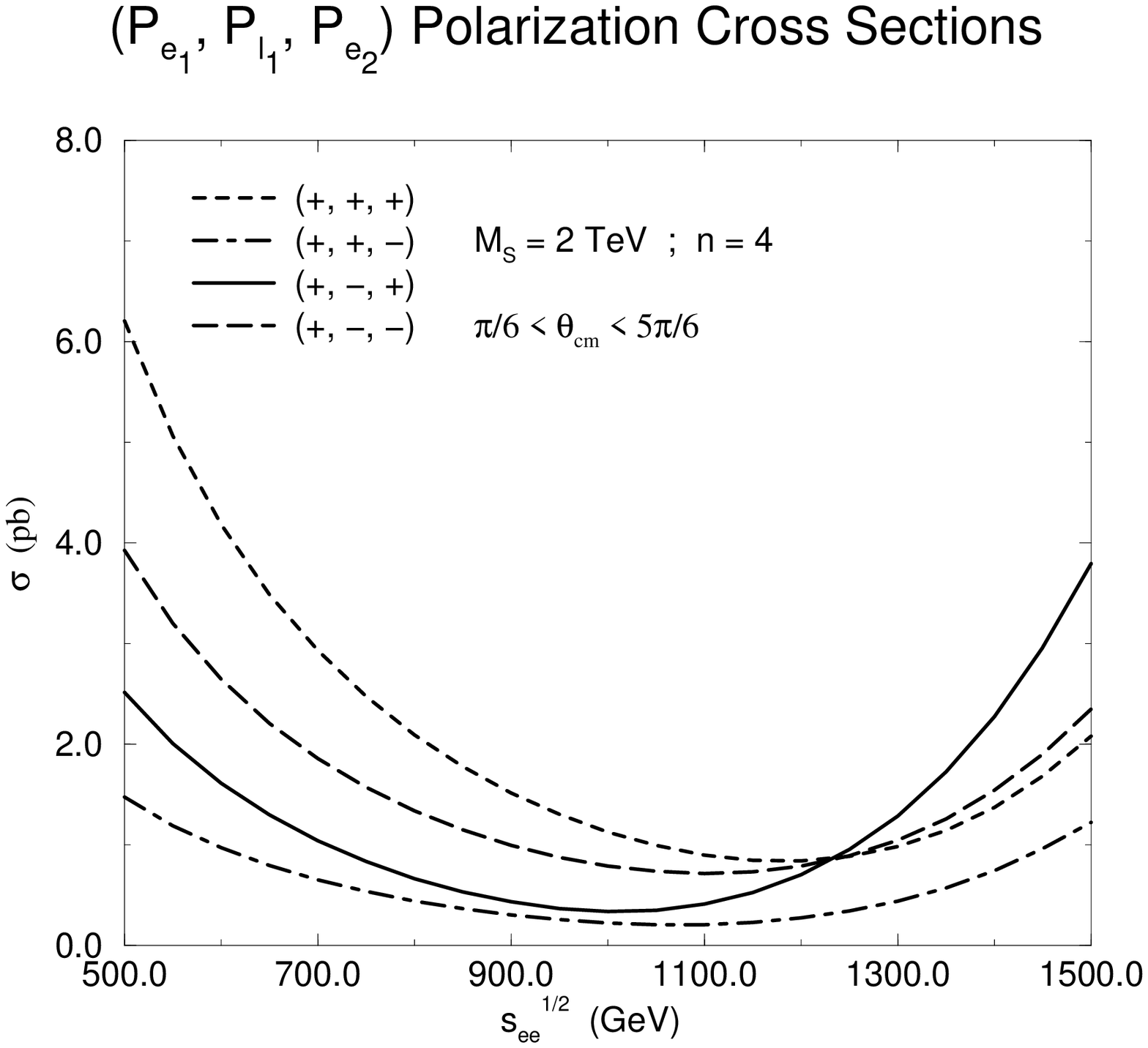}} 
\fcaption{SM + WSQG cross sections with four independent initial electron and 
laser beam polarizations.  Here, 
$M_S = 2$ TeV and $n = 4$ ($\gege$).} 
\label{polsigs} 
\end{figure}

\begin{figure}[htbp]    
\centerline{\epsfxsize=10truecm \epsfbox{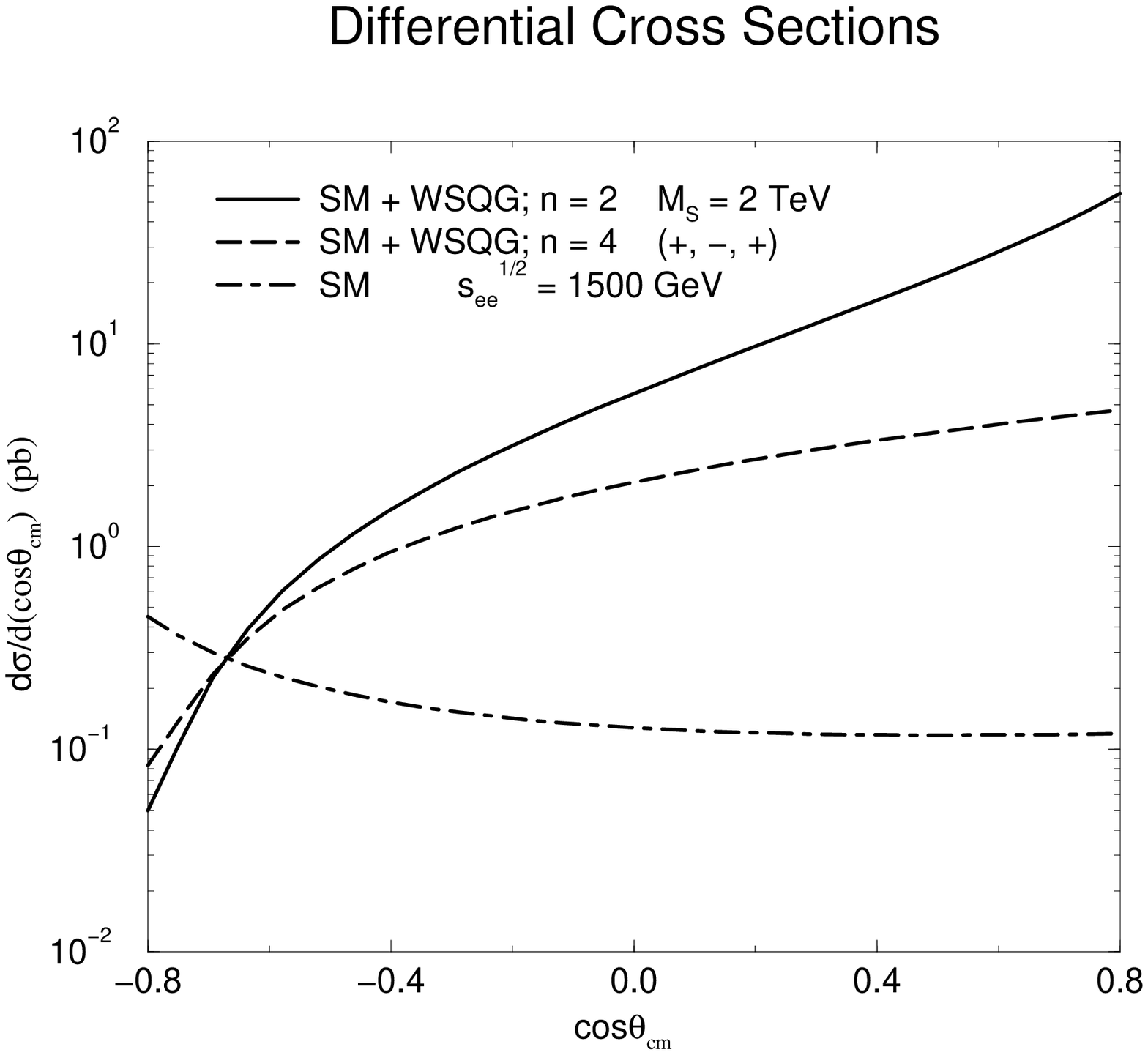}}
\fcaption{SM + WSQG and SM differential cross sections at $\sqrt{s_{ee}} = 1500$ GeV for 
the $(+, -, +)$ polarization.  
Here, $M_S = 2$ TeV and $n = 2, 4$, for the WSQG contributions ($\gege$).}
\label{dcs}
\end{figure}

The cross sections for $w = - 1$ are larger than the ones for $w = +1$, as evident from Eq.  (\ref{M+j2}).  However, we note that it is
more conservative to choose $w = + 1$, in order to avoid an overestimate of the effects, and in any case, this is the choice that
follows from a straightforward use of the low energy effective Lagrangian.  Nonetheless, in the following, we will present results
indicating that the discovery reach of the NLC for the value of the parameter $M_S$ is approximately the same for $w = \pm 1$.  Fig. 
(\ref{polsigs}) shows the  effect of polarization on the cross section, where we have chosen $M_S = 2$ TeV and $n = 4$.  We see that
the polarization choice $(P_{e_1}, P_{l_1} , P_{e_2}) = (+, -, +)$ gives the dominant cross section at high energies.  The differential
cross sections with polarization $(+, -, +)$ at $\sqrt{s_{ee}} = 1500$ GeV for SM, and SM + WSQG, with $M_S = 2$ TeV and $n = 2,
4$, are presented in Fig. (\ref{dcs}).  We see that at this value of $\sqrt{s_{ee}}$, due to spin-2 KK graviton exchange, the SM + WSQG angular distributions for $\gege$
are very different from the prediction of the SM.  The SM + WSQG differential cross section with $n = 2$ is enhanced in the forward
direction, since $\ln (M_S^2/{\hat t}) \to \infty$ as $\theta_{cm} \to 0$.

\begin{figure}[htbp]    
\centerline{\epsfxsize=10truecm \epsfbox{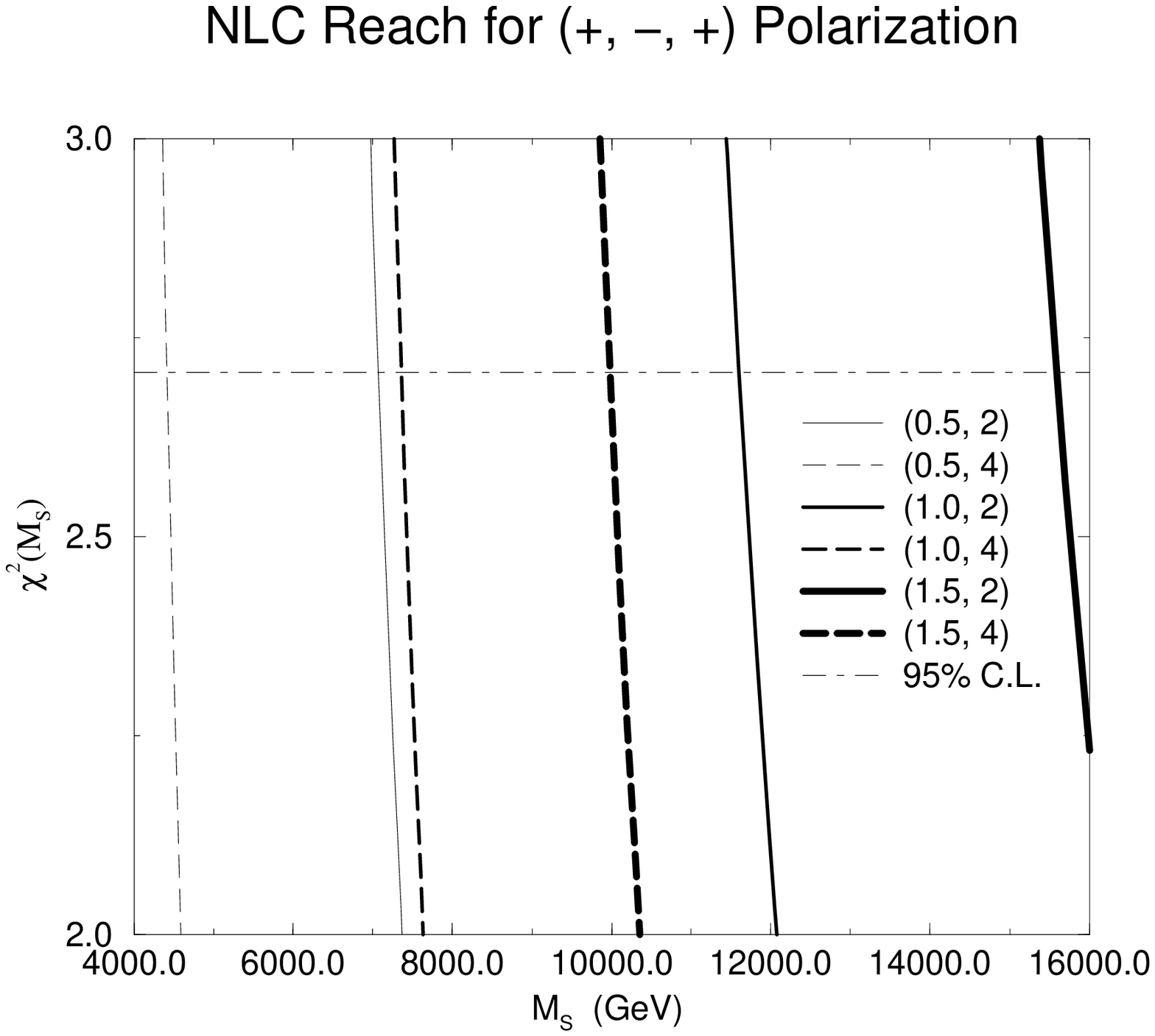}}
\fcaption{The solid and the dashed lines represent the $\chi^2$ as a function 
of $M_S$ for the cases 
$n = 2$ and $n = 4$, respectively, at three values of $\sqrt{s_{ee}}$ with polarization $(+, -, +)$.  
The numbers in the parentheses denote the value of 
$\sqrt{s_{ee}}$, in TeV, and $n$, respectively.  The dot-dashed line marks the 
reach at the $95\%$ confidence level ($\gege$).}
\label{reach}
\end{figure}

\begin{figure}[htbp]    
\centerline{\epsfxsize=10truecm \epsfbox{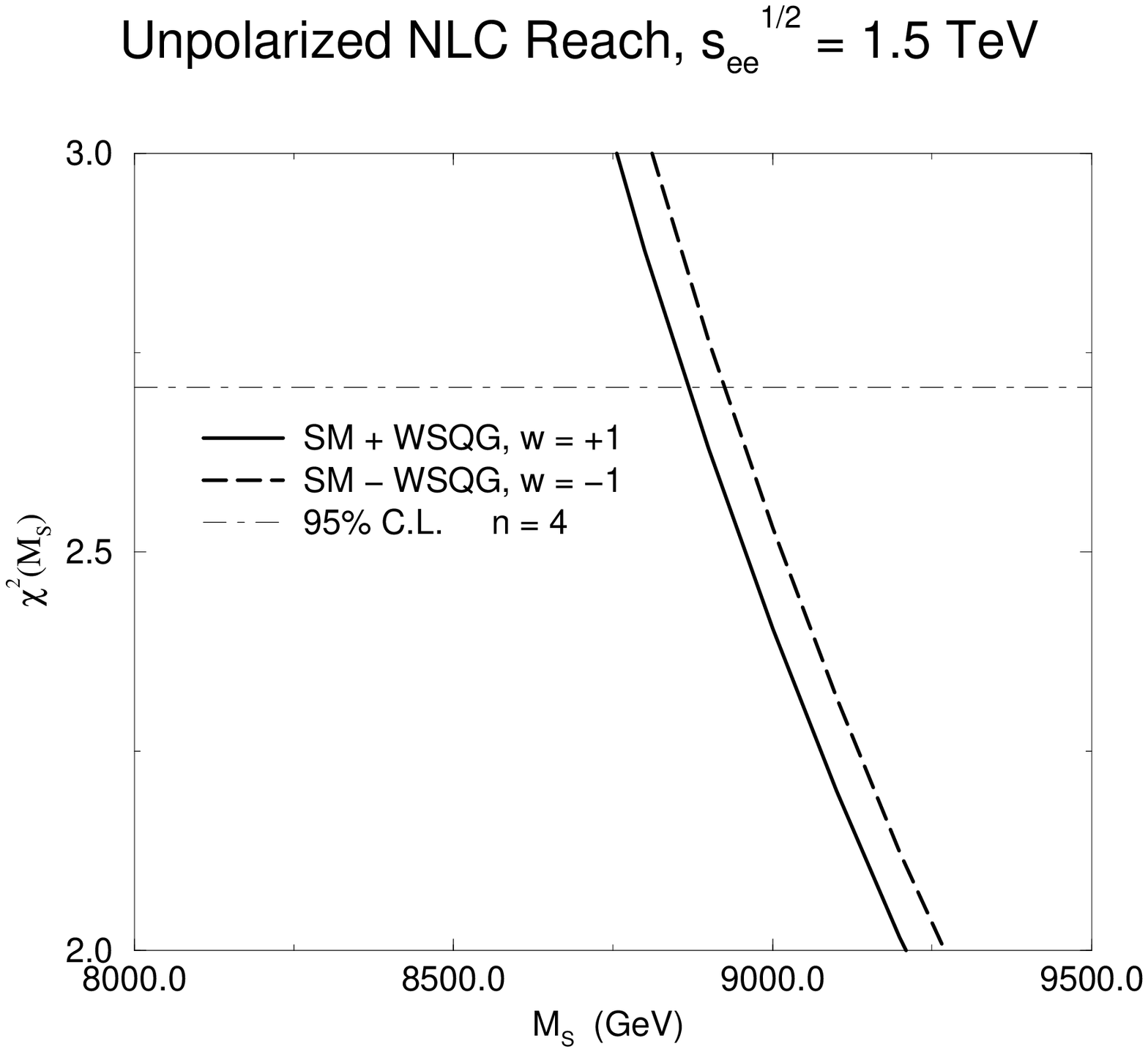}}
\fcaption{The solid and the dashed lines, corresponding to $w = \pm 1$, 
respectively, represent the $\chi^2$ as a 
function of $M_S$ for unpolarized beams as a function of $M_S$, with $n = 4$, at
 $\sqrt{s_{ee}} = 1500$ GeV.  
The dot-dashed line marks the reach at the $95\%$ confidence level ($\gege$).}
\label{unpreach}
\end{figure}

The $M_S$ reach at the NLC with center of mass energies of 500 GeV, 1000 GeV, and 1500 GeV, for the $(+, -, +)$ polarization
choice, are shown in Fig. (\ref{reach}).  The smallest reach in Fig. (\ref{reach}) is about 4 TeV for $n = 4$ and $\sqrt{s_{ee}} = 500$
GeV and the largest reach is a bout 16 TeV for $n = 2$ and $\sqrt{s_{ee}} = 1500$ GeV.  Note that the reach for $n = 2$ at
$\sqrt{s_{ee}} = 500$ GeV is about 7 TeV or approximately $14 \sqrt{s_{ee}}$.  According to Eq.  (\ref{chi2}), the reach can be
improved by increasing the luminosity $L$.  However, we have checked that using $L = 200$ fb$^{-1}$ per year does not improve
the reach significantly.  We present the unpolarized NLC reach for $n = 4$ and $w = \pm 1$ at $\sqrt{s_{ee}} = 1500$ GeV in Fig. 
(\ref{unpreach}).  We see that the effects of the sign of $w$ on the reach are not significant.  Comparing the curve marked $(1.5, 4)$
in Fig. (\ref{reach}) with the curve for $w = +1$ in Fig.  (\ref{unpreach}) shows that the reach is enhanced with the use of the $(+, -,
+)$, since the $(+, -, +)$ back-scattered $\gamma$-beam has a larger number of hard photons than the 
unpolarized beam.\cite{Ginz2}

\section{\boldmath$\gggg$ at a \boldmath$\gamma \gamma$ collider}

We consider the process $\gamma (k_1) \gamma (k_2) \to \gamma (p_1) \gamma (p_2) $, where $k_1$ and $k_2$ are the initial and
$p_1$ and $p_2$ are the final 4-momenta of the photons.  This process has the advantage that it receives contributions from the 
SM only at the loop level and, therefore, could in principle be sensitive to new physics at the tree level.  We define
$s \equiv (k_ 1 + k_2)^2, t \equiv (k_1 - p_1)^2$, and $u \equiv (k_1 - p_2)^2$.  Helicity amplitudes are denoted by $M_{i j k l}$,
where $i, j, k, l = \pm$, and $(i, j)$ are the helicities of the $(k_1, k_2)$ photons, and $(k, l)$ are the helicities of the $(p_1, p_2)$
photons.  The 1-loop helicity amplitudes of the SM are in general complicated.  However, in the limit $s, |t|, |u| \gg m^2$, where $m$
is the mass of a $W$ boson, a quark, or a charged lepton, these amplitudes can be approximated by those parts of them that receive
logarithmic enhancements\cite{Gounaris}.  Except for the contribution of the top quark loop which does not affect our results
significantly\cite{Gounaris}, these leading amplitudes provide a good approximation at the energies of the NLC in the $\gamma
\gamma$ collider mode.  Each high energy $\gamma$-beam can be achieved by the back scattering of a laser beam from an
$e$-beam, as was discussed in the previous section.  WSQG contributes to $\gggg$ through the exchange of towers of KK
gravitons in the $s$-, $t$-, and $u$-channels at the leading order.  The contents of this section have some overlap with the results
of Ref.~[\citenum{KC}].  

For various choices of the pairs $(P_{e_1}, P_{l_1})$ and
$(P_{e_2}, P_{l_2})$ of the the two beams, the differential cross section 
$d \sigma/d \Omega$ is given by
\[
\frac{d \sigma}{d \Omega} = \frac{1}{128 \, \pi^2 \, s_{ee}} \int \int d x_1 d x
_2 \left[\frac{f(x_1) \, f(x_2)}{x_1 \, x_2}\right]
\]
\begin{equation}
\times
\left[\left(\frac{1 + \xi_2(x_1) \, \xi_2(x_2)}{2}\right)|M_{++}|^2 + 
\left(\frac{1 - \xi_2(x_1) \, \xi_2(x_2)}{2}\right)|M_{+-}|^2\right],
\label{ggggdiffcs}
\end{equation}
where $x_1$ and $x_2$ are the energy fractions for the two beams, given by 
Eq. (\ref{x}).  Different choices of $(P_{e_1}, P_{l_1})$ and $(P_{e_2}, P_{l_2})$ in $(f(x_1), 
\xi_2(x_1))$ and $(f(x_2), \xi_2(x_2))$, respectively, yield different polarization cross sections.

The logarithmically
enhanced SM amplitudes,used here, are valid when $s, |t|, |u| \gg m_W^2$.  However, we see that to have a good approximation,
we must demand ${\hat s}, |{\hat t}|, |{\hat u}| \gg m_W^2$.  To avoid restricting the phase space too much, and in
order to have a good approximation to the SM amplitudes, we will impose the cuts $\theta_{cm} \in [\pi/6, 5 \pi/6]$, 
$x_1 \in [\sqrt{0.4}, x_{1 max}]$, and $x_2 \in [\sqrt{0.4}, x_{2 max}]$, 
where $x_{1 max} = x_{2 max} = (z)/(1 + z)$.  These cuts ensure that the integrations are always performed in a 
region where ${\hat s}, |{\hat t }|, |{\hat u}| > m_W^2$.  

\begin{figure}[htbp]    
\centerline{\epsfxsize=10truecm \epsfbox{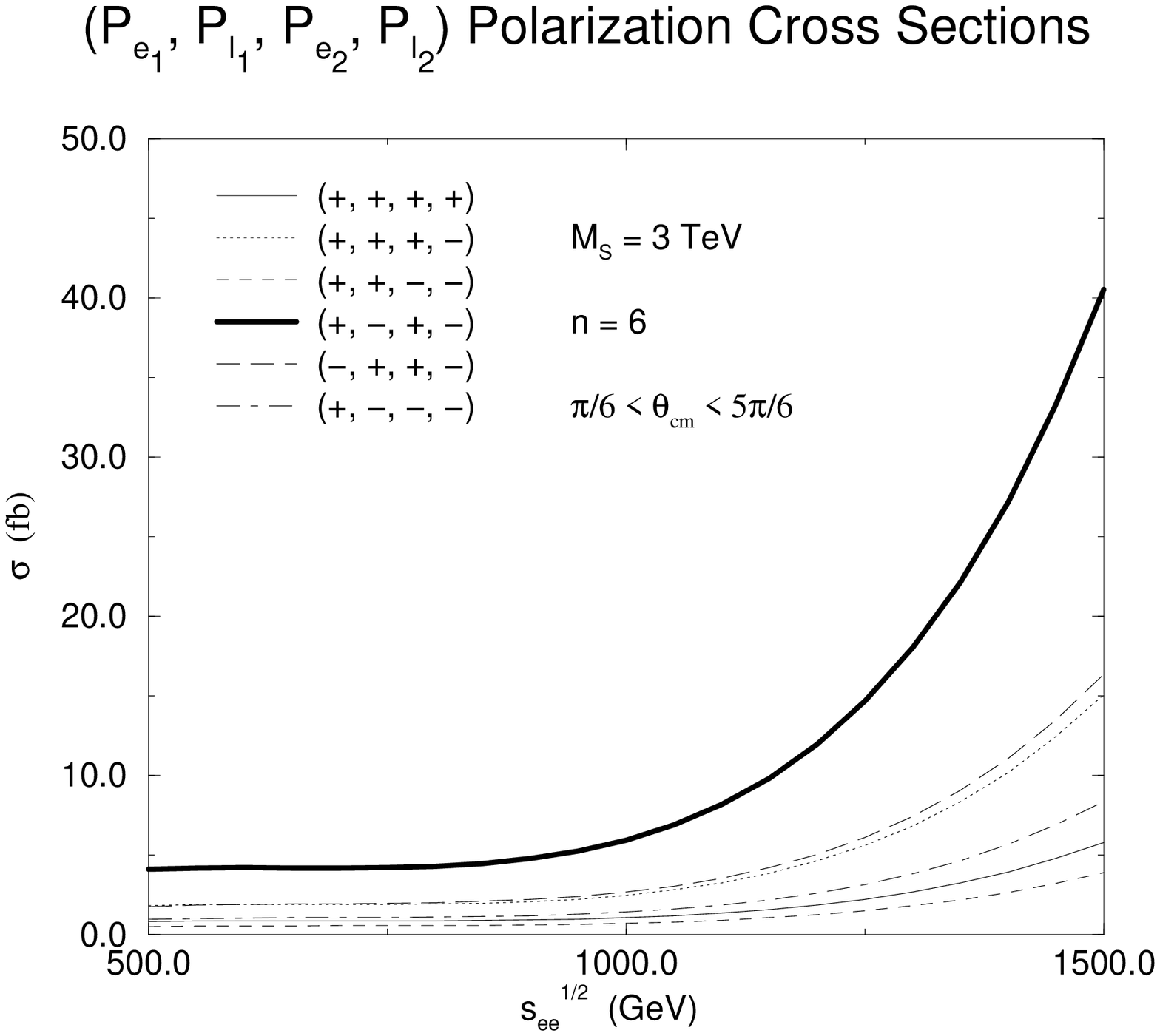}}
\fcaption{SM + WSQG cross sections for six independent initial electron and 
laser beam polarizations.  Here, 
$M_S = 3$ TeV and $n = 6$ ($\gggg$).}
\label{sixpols}
\end{figure}

\begin{figure}[htbp]    
\centerline{\epsfxsize=10truecm \epsfbox{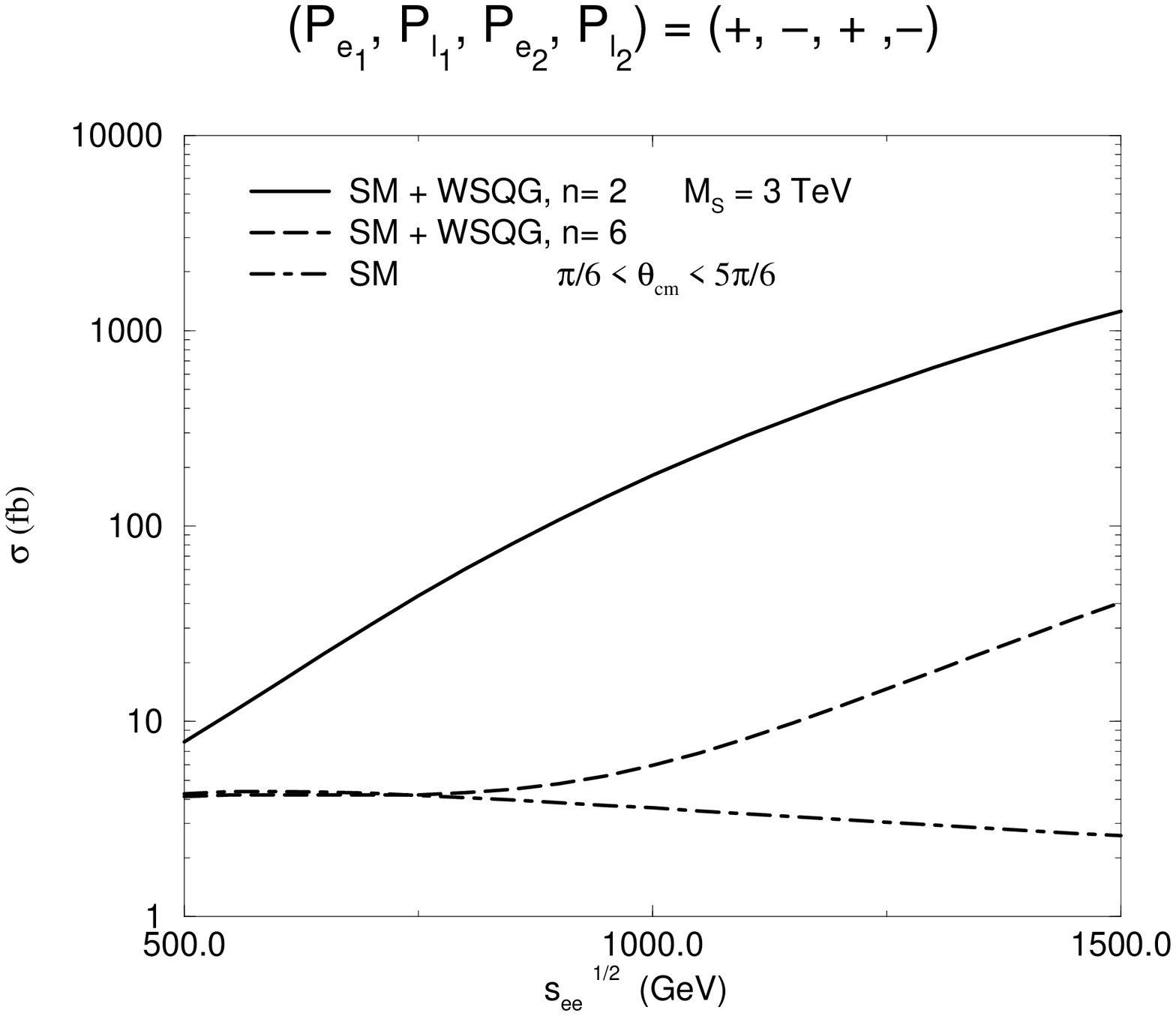}}
\fcaption{SM + WSQG and SM cross sections for the $(+, -, +, -)$ polarization.  Here, 
$M_S = 3$ TeV and $n = 2, 6$, for the WSQG contributions ($\gggg$).}
\label{26sm}
\end{figure}

\begin{figure}[htbp]    
\centerline{\epsfxsize=10truecm \epsfbox{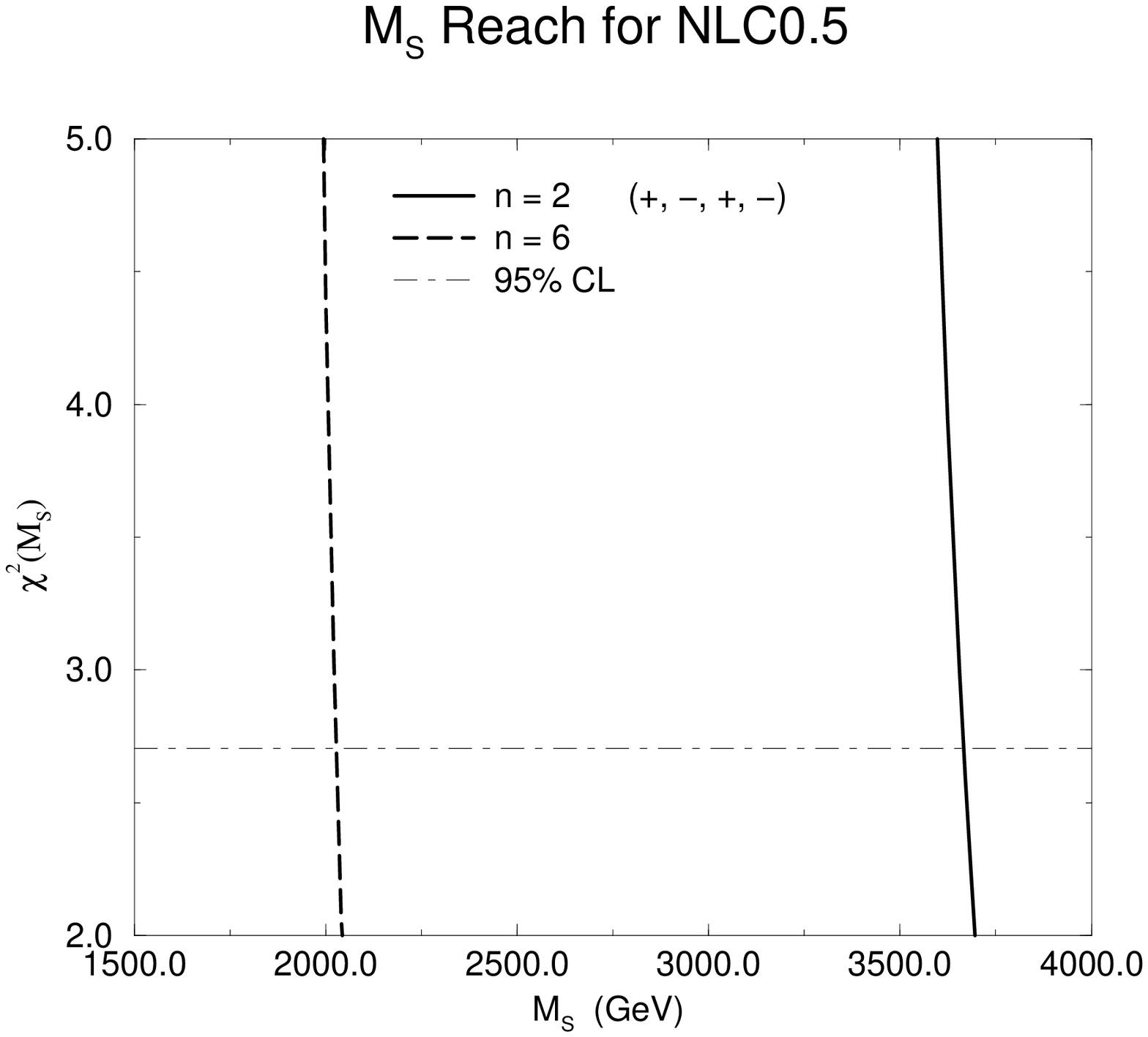}}
\fcaption{The $M_S$ reach for NLC0.5.  The solid and the dashed lines represent 
the $\chi^2$ as a function of $M_S$ for the cases 
$n = 2$ and $n = 6$, respectively.  The dot-dashed line marks the reach at the $
95\%$ confidence level ($\gggg$).}
\label{reach0.5}
\end{figure}

\begin{figure}[htbp]    
\centerline{\epsfxsize=10truecm \epsfbox{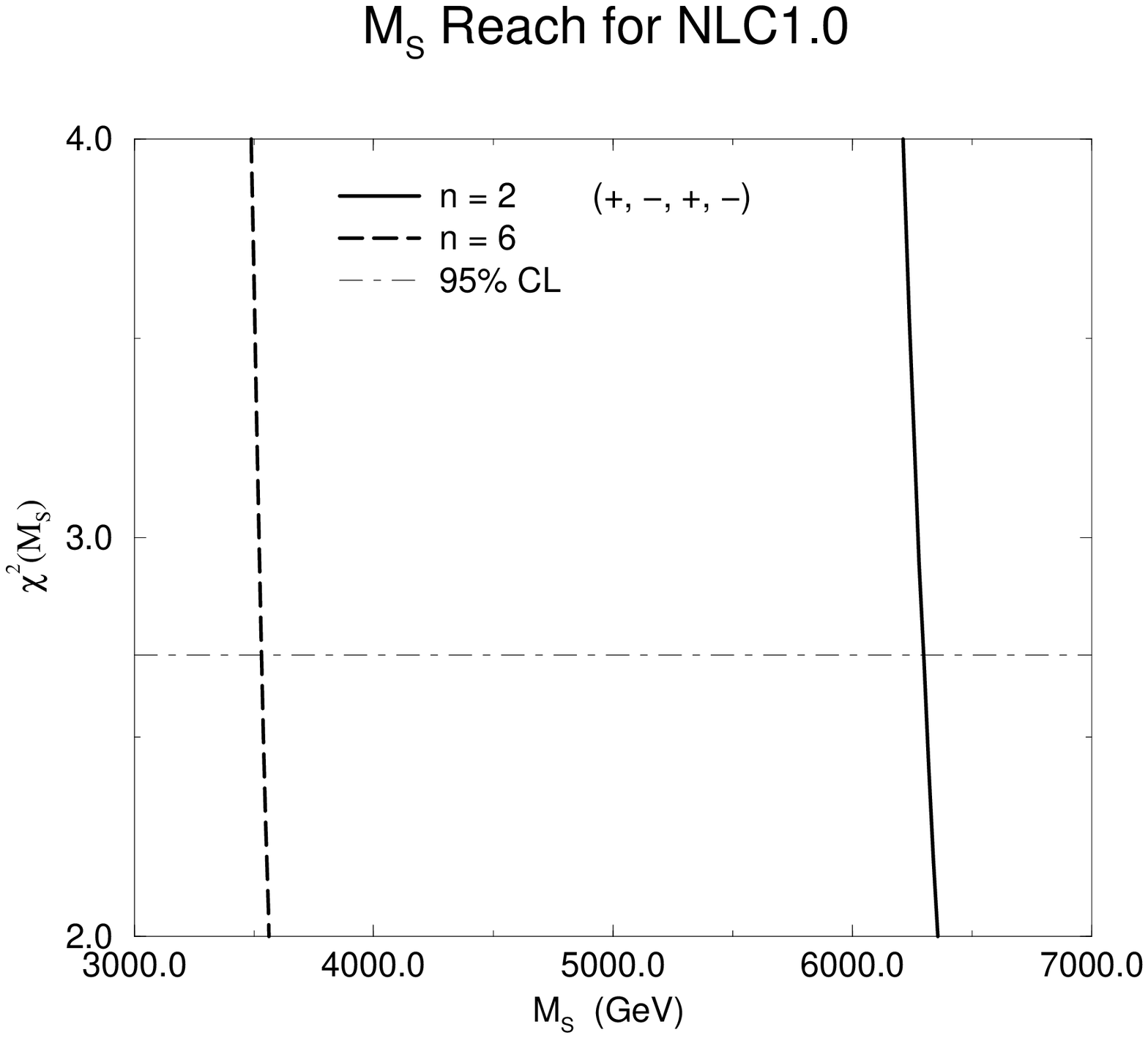}}
\fcaption{The $M_S$ reach for NLC1.0.  The solid and the dashed lines represent 
the $\chi^2$ as a function of $M_S$ for the cases 
$n = 2$ and $n = 6$, respectively.  The dot-dashed line marks the reach at the $
95\%$ confidence level ($\gggg$).}
\label{reach1.0}
\end{figure}

\begin{figure}[htbp]    
\centerline{\epsfxsize=10truecm \epsfbox{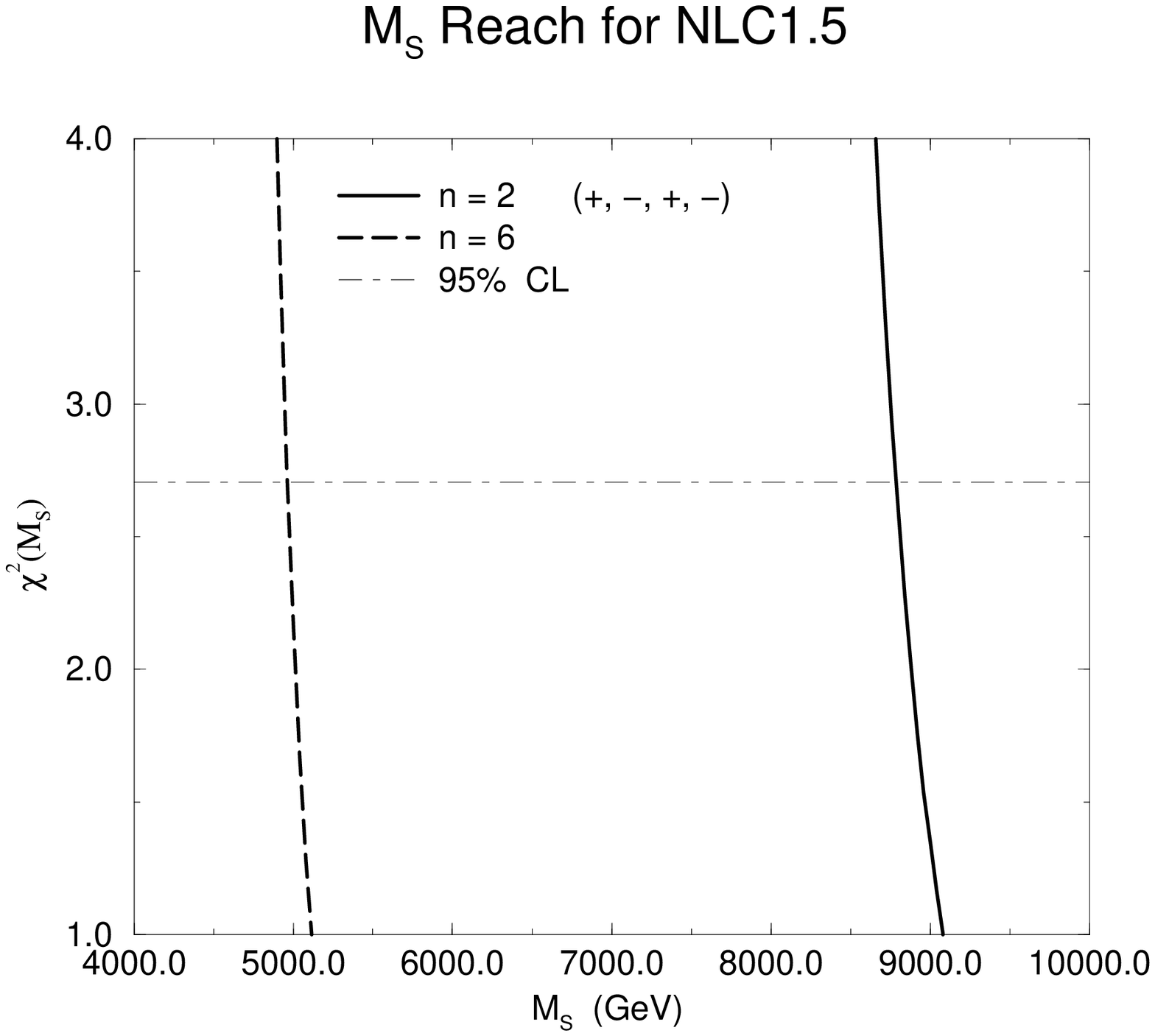}}
\fcaption{The $M_S$ reach for NLC1.5.  The solid and the dashed lines represent 
the $\chi^2$ as a function of $M_S$ for the cases 
$n = 2$ and $n = 6$, respectively.  The dot-dashed line marks the reach at the $
95\%$ confidence level ($\gggg$).}
\label{reach1.5}
\end{figure}

The results that are presented for $\gggg$ here correspond to the choice $w = +1$.  The six SM + WSQG cross sections, for $M_S
= 3$ TeV and $n = 6$, in Fig. (\ref{sixpols}), correspond to six independent choices for the polarizations $(P_{e_1}, P_{l_1},
P_{e_2}, P_{l_2})$ of the electron and the laser beams of the photon collider.  These cross sections are plotted
versus the center of mass energy of the $e$-beams, $\sqrt{s_{ee}}$.  The curves in this figure show a sensitive dependence on the
choices of the polarizations for $\sqrt{s_{ee }} \grtsim 1$ TeV, with the $(+, -, +, -)$ polarization giving the largest cross section at
high energies.  In Fig. (\ref{26sm}), choosing $M_S = 3$ TeV and $n =2, 6$, we compare the SM + WSQG cross sections with that
of the SM in the typical proposed NLC center of mass energy range $\sqrt{s_{ee}} \in [500, 1500]$ GeV.  We have chosen the $(+,
-, +, -)$ polarization for all three curves, since this choice yields the largest gravity cross section, as shown in Fig. (\ref{sixpols}). 
The plots in Figs. (\ref{reach0.5}), (\ref{reach1.0}), and (\ref{reach1.5}) show the $95\%$ confidence level experimental reach for
$M_S$ at NLC0.5, NLC1.0, and NLC1.5, respectively.  The lowest reach in $M_S $ is
about 2 TeV for $n = 6$ at NLC0.5 and the largest $M_S$ reach is about 9 TeV for $n = 2$ at NLC1.5.  These values are obtained
for $L = 100$ fb$^{-1}$ per year.

\section{Concluding Remarks}

In the above, we showed that given $\sqrt{s_{ee}} \sim 1$ TeV, and $L \sim 100$ fb$^{-1}$ per
year at the NLC with the photon collider option, WSQG can be probed over the interesting range 
1 TeV $\lsim M_S \lsim$ 10 TeV, by studying $\gege$ and $\gggg$.  It was demonstrated that beam polarization plays an important role in optimizing the discovery reach for the signatures of WSQG.  Since $e^-$-beams can be polarized much more efficiently than 
$e^+$-beams, measurements of the signatures of WSQG in the channels discussed here can be best achieved at an $e^- e^-$ 
collider.

\newpage

\nonumsection{References}

\end{document}